\documentclass[11pt]{article}
\usepackage{amsmath}
\usepackage{amsfonts}
\usepackage[dvips]{epsfig}

 \advance \textwidth by -2in
 \advance \textheight by -3in
\def\##1{{\underline #1}}
\def\=#1{\underline{\underline{#1}}}

\def\+#1{\underline{\underline #1}}
\def\*#1{\underline{\underline{\bf #1}}}

\def\eps{\epsilon}
\def\epso{\epsilon_0}
\def\muo{\mu_0}
\def\ko{k_0}

\def\.{\mbox{ \tiny{$^\bullet$} }}

\def\le{\left(}
\def\ri{\right)}
\def\les{\left[}
\def\ris{\right]}

\def\c#1{\cite{#1}}
\def\l#1{\label{#1}}
\def\r#1{(\ref{#1})}

\begin{document}

\noindent {\bf NEGATIVE REFLECTION IN A FARADAY CHIRAL MEDIUM
  }
\vskip 0.2cm

\noindent  {\bf Tom G. Mackay$^1$ and Akhlesh Lakhtakia$^2$} \vskip
0.2cm

\noindent {\sf $^1$ School of Mathematics\\
\noindent University of Edinburgh\\
\noindent Edinburgh EH9 3JZ, United Kingdom} \vskip 0.4cm

\noindent {\sf $^2$ CATMAS~---~Computational \& Theoretical Materials Sciences Group \\
\noindent Department of Engineering Science \& Mechanics\\
\noindent 212 Earth \& Engineering Sciences Building\\
\noindent Pennsylvania State University, University Park, PA
16802--6812} \vskip 0.4cm

\noindent {\bf ABSTRACT:}  The  four wavenumbers associated with
planewave propagation in a Faraday chiral medium (FCM) with
relatively huge magnetoelectric coupling  give rise to enhanced
possibilities for negative--phase--velocity propagation and
therefore negative refraction. They can also give rise to the
phenomenon of negative reflection. In particular, for a
nondissipative example,  we deduced that an incident plane wave with
positive/negative phase velocity can result in a negatively
reflected plane wave with negative/positive phase velocity, as well
as a positively reflected plane wave with positive/negative phase
velocity.


\vskip 0.4cm

\vspace{10mm}

\noindent{\bf 1. INTRODUCTION}

The scope for exotic~---~and potentially useful~---~electromagnetic
phenomenons to arise is greatly extended in
 anisotropic mediums, and especially
 bianisotropic mediums, as compared with isotropic mediums \c{mac05,PiO}.
This letter concerns a particular bianisotropic medium called a
Faraday chiral medium (FCM). FCMs combine natural optical
activity (as exhibited by isotropic chiral mediums
\c{Beltrami}) with Faraday rotation (as exhibited by
gyrotropic mediums \c{Lax,Collin}). They  may be
  theoretically conceptualized as
homogenized composite mediums arising  from the blending together of
isotropic chiral mediums  with either  magnetically biased ferrites
\c{WLM98} or magnetically biased plasmas \c{WM00}. Accordingly, a
wide range of constitutive parameters for FCMs may be envisaged.

In a recent study, we identified the propensity of FCMs to support
planewave propagation with negative phase velocity, particularly
when the magnetoelectric constitutive parameters are relative large
\c{ML_PRE04}. Similar behavior in isotropic chiral mediums and their
nonreciprocal counterparts can be traced back to 1986 \c{LVV86},
but is only nowadays being considered carefully
\c{TGM_MOTL, BSPKA}, following the successful exhibition of the
phenomenon of \emph{negative refraction} by isotropic
achiral materials \c{SAR}. In the following sections we consider the phenomenon
of \emph{negative reflection} in FCMs. Negative reflection has previously
been described for isotropic chiral mediums \c{ZC,FN} and anisotropic
dielectric mediums \c{Wang05}, but not to our knowledge for
bianisotropic mediums.

In the following sections, an $\exp(-i \omega t)$ time dependence is
assumed;  the permittivity and permeability of free space are
denoted by $\epso$ and $\muo$, respectively; 3--vectors are
underlined, with
  the caret ($\,\hat{}\,$) denoting a unit vector; and
  3$\times$3 dyadics are double underlined.

 \noindent{\bf 2. PLANEWAVE
PROPAGATION}

In preparation for solving a boundary--value problem in Section 3,
let us consider planewave propagation
in a FCM
   characterized
  by  the frequency--domain constitutive relations
   \c{Engheta,WL98}
\begin{equation}
\left.
\begin{array}{l}
  \#D (\#r) = \=\eps\.\#E (\#r) + \=\xi\.\#H (\#r) \,\\[5pt]
  \#B (\#r) = - \=\xi\.\#E (\#r) + \=\mu\.\#H (\#r) \, \l{FCM_cr}
\end{array}
\right\},
\end{equation}
with constitutive dyadics
\begin{equation}
\left.
\begin{array}{l}
  \=\eps =
\epso \les \, \eps \, \=I - i \eps_g \, \hat{\#u} \times \=I +
\le \, \eps_z - \eps \, \ri \,  \hat{\#u}\,  \hat{\#u} \, \ris\\
\vspace{-2mm} \\
  \=\xi =
i \sqrt{\epso \muo} \, \les \, \xi \, \=I - i \xi_g \, \hat{\#u}
\times \=I + \le \, \xi_z - \xi \, \ri \,
  \hat{\#u} \,  \hat{\#u}
\ris\\
\vspace{-2mm} \\
  \=\mu =
\muo \les \, \mu \, \=I - i \mu_g \, \hat{\#u} \times \=I + \le \,
\mu_z - \mu \, \ri \,  \hat{\#u}\,  \hat{\#u} \, \ris
\end{array}
\right\}. \l{FCM}
\end{equation}
Simplifying for the sole purpose of exemplifying negative reflection, we choose the  distinguished axis of the FCM to
coincide with the $z$ axis; i.e., $\hat{\#u} = \hat{\#z}$.

Let us also confine ourselves to propagation in the $xz$ plane. Plane waves with field phasors
\begin{equation}
\left.\begin{array}{l}
\#E (\#r) = A\,\#E_{\,0}\, \exp \le i \#k\.\#r   \ri\\[5pt]
\#H (\#r) = A\,\#H_{\,0} \, \exp\le i \#k\.\#r   \ri
\end{array}\right\}
\l{pw}
\end{equation}
can propagate in the FCM, where $A$ is a complex--valued amplitude
and the wave vector
\begin{equation}
\#k = \kappa \, \hat{\#x} + \sqrt{k^2 - \kappa^2} \, \hat{\#z} =
k\,\hat{\#k} =k(\hat{\#x}\sin\psi+\hat{\#z}\cos\psi)
\end{equation}
contains the real--valued quantity $\kappa$ that is fixed by the
incidence conditions in Section 3. The angle $\psi$ is defined in
the clockwise sense with respect to the $+z$ axis in the $xz$ plane.

In order to determine $\#k$, $\#E_{\,0}$, and $\#H_{\,0}$, we
substitute \r{pw} in the Maxwell curl postulates $\nabla\times\#E
(\#r)=i\omega\#B(\#r)$ and $\nabla\times\#H (\#r)=-i\omega\#D(\#r)$,
and then use the constitutive relations of the FCM. In general, we
find four independent wavevectors for a specific
$\kappa$~---~namely,
\begin{equation}
\#k_j = \kappa \, \hat{\#x} + \sqrt{k_j^2 -
\kappa^2}\,\hat{\#z}=k_j(\hat{\#x}\sin\psi_j+\hat{\#z}\cos\psi_j)
\,,\qquad j\in\les 1,4\ris, \label{def-psi}
\end{equation}
Corresponding to each $\#k_j$, there exist ${\#E}_{\,0j}$ and $
{\#H}_{\,0j}= \=\mu^{-1} \. \les (1/\omega )\le \#k_j \times \=I
\,\ri + \=\xi \ris \. \#E_{\,0j}$. Thus, for the purposes of Section
3, the complete representation of plane waves in the FCM for a
specified $\kappa$ is
\begin{equation}
\left.
\begin{array}{l}
\#E(\#r) = \sum_{j=1}^{4}\, A_j \,\#E_{\,0j}\, \exp \le i \#k_j\.\#r\ri\\[5pt]
\#H(\#r) = \sum_{j=1}^{4}\, A_j \,\#H_{\,0j}\, \exp \le i
\#k_j\.\#r\ri
\end{array}
\right\}\,, \l{FCMrep}
\end{equation}
wherein we have neglected the remote possibility of the existence of
Voigt waves \c{Gerardin}.

The associated time--averaged Poynting vector is given by
\begin{equation}
\#P(\#r)=  \sum_{j=1}^{4}\, \#P_j (\#r) = \frac{1}{2}
\sum_{j=1}^{4}\,\vert A_j\vert^2\, \mbox{Re} \, \le \, \#E_{\,0j}
\times \#H_{\,0j}^* \,\ri\,
 \exp\les-2\,
\mbox{Im}\le\#k_j\ri\.\#r\ris \,.
\end{equation}
The phase velocity of
the  j$^{th}$ plane wave in \r{FCMrep} is \emph{positive}
 if $\mbox{Re}\le \#k_j   \ri\. \#P_j(\#r) >
0$ and \emph{negative}  if $\mbox{Re}\le \#k_j  \ri \. \#P_j(\#r) < 0$.

\noindent{\bf 3. BOUNDARY--VALUE PROBLEM}

Suppose that the half--space $z > 0$ is filled  with a FCM described
by the constitutive relations \r{FCM_cr},  whereas  the half--space
$z<0$ is vacuous. A distant source in the FCM--filled half--space
supposedly launches a plane wave towards the boundary $z=0$. The
wave vector of this plane wave lies wholly in the $xz$ plane, and
its projection on the boundary $z=0$ is denoted by $\kappa$.

Equation \r{FCMrep} holds in the region $z\geq0$ with the following
stipulations:
\begin{itemize}
\item[(i)] The indexes $j=1$ and $j=2$ are reserved for plane waves
that transport energy towards the boundary $z=0$, i.e.,
\begin{equation}
\hat{\#z}  \. \#P_j(\#r)<0\,,\quad j\in\les 1,2\ris\,.
\end{equation}
These two indexes thus identify
\emph{incident} plane waves.
\item[(ii)] The indexes $j=3$ and $j=4$ are reserved for plane waves
that transport energy away from the boundary $z=0$, i.e.,
\begin{equation}
\hat{\#z}  \. \#P_j(\#r)>0\,,\quad j\in\les 3,4\ris\,.
\end{equation}
These two indexes thus identify
\emph{reflected} plane waves.
\item[(iii)] Either $A_1=0$ or $A_2=0$, but not both. When $A_1\ne0$,
$\psi_1$ can be called as the angle of incidence; when $A_2\ne0$,
$\psi_2$ is the angle of incidence.
\end{itemize}

If  wavevectors of the  incident and  a reflected plane wave are oriented in
either (i) the same quadrant of the $xz$ plane or (ii)  diagonally
opposite quadrants of the $xz$ plane, then the reflection is called
negative; otherwise, the reflection is called positive.

In the vacuous region, the transmitted field phasors are given by
\begin{equation}
\left.
\begin{array}{l}
\#E(\#r) = (B_s \hat{\#y} + B_p \hat{\#p}) \exp\les i (\kappa x - \alpha z)\ris
\\[5pt]
\#H(\#r) = \eta_0^{-1}\,(B_s \hat{\#p} - B_p \hat{\#y}) \exp\les i (\kappa x - \alpha z)\ris
\end{array}\right\}\,,\quad z \leq 0\,,
\end{equation}
where $B_s$ and $B_p$ are unknown amplitudes for the $s$ and $p$
polarization states, and
\begin{equation}
\left.
\begin{array}{l}
\hat{\#p}=(\alpha \hat{\#x}+\kappa\hat{\#z})/k_0
\\[5pt]
\alpha = +\sqrt{k_0^2-\kappa^2}\\[5pt]
k_0=\omega\sqrt{\eps_0\mu_0}\\[5pt]
\eta_0=\sqrt{\mu_0/\eps_0}
\end{array}\right\}\,.
\end{equation}

Imposition of the usual boundary conditions across the boundary $z=0$  yields
four algebraic equations:
\begin{eqnarray}
\l{ae1}&&\hat{\#x}\.\sum_{j=1}^{4}\, A_j \,\#E_{0j} = B_p
\,\alpha/k_0\, \,,\\ \l{ae2}&&\hat{\#y}\.\sum_{j=1}^{4}\, A_j
\,\#E_{0j} = B_s
\,,\\
&&\hat{\#x}\.\sum_{j=1}^{4}\, A_j \,\#H_{0j} =B_s \, \alpha/ \le k_0
\eta_0 \ri\,
\,,\\
&&\hat{\#y}\.\sum_{j=1}^{4}\, A_j \,\#H_{0j} =- B_p/\eta_0 \,.
\l{ae4}\end{eqnarray} We recall that either $A_1=0$ or $A_2=0$, but
not both, for our present purpose.

\noindent{\bf 4. A NUMERICAL ILLUSTRATION}

In order to provide a numerical illustration, we considered
  the nondissipative regime wherein $\rho$, $\rho_g$, $\rho_z \in
\mathbb{R}$ ($\rho= \eps, \xi, \mu)$, and $k_j \in
\mathbb{R}\;\forall j\in[1,4]$. For the results presented
in this section, we fixed $\eps = 3 $, $\eps_g =
1$, $\eps_z = 2.5  $; $\xi = 10$, $\xi_g = 6 $, $\xi_z = 7  $; $\mu
= 2 $, $\mu_g = 0.5 $ and $\mu_z = 1.5 $.

\begin{figure}[!ht]
\centering \psfull  \epsfig{file=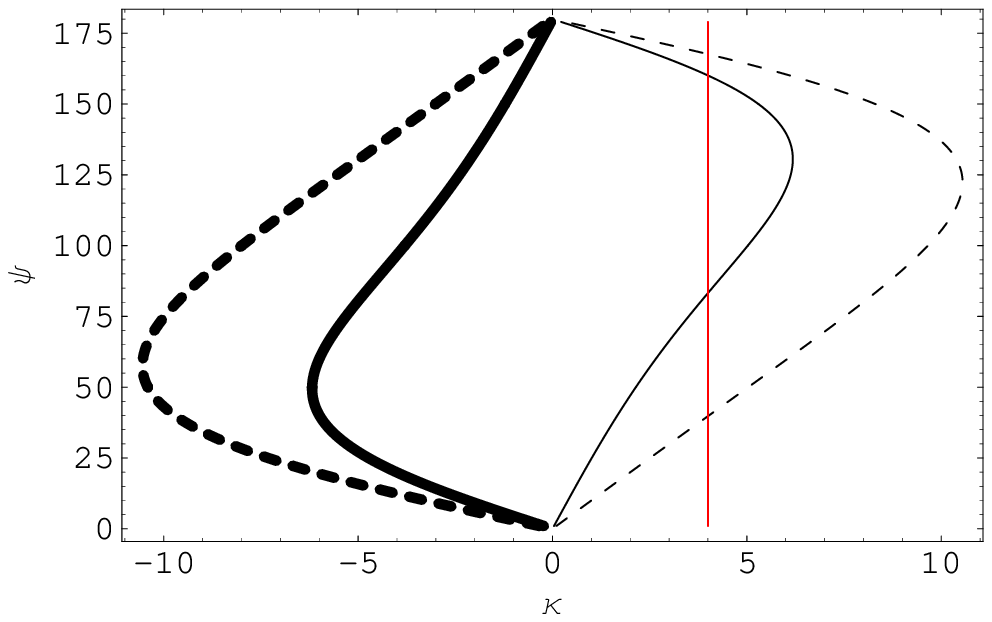,width=5.0in}
  \caption{\label{fig1} The wavevector orientation angle $\psi$
(in degree) plotted against  $\kappa$ (normalized with respect to
$\ko$). There are four possible values of $\psi$ at each value of
$\kappa$. The vertical line indicates $\kappa = 4 \ko$. }
\end{figure}

The  values  of $\psi$ for each value of $\kappa$ are plotted in
Figure~\ref{fig1}. Let us focus attention on a specific value of
$\kappa$; e.g., $\kappa = 4 \ko$. As highlighted in
Figure~\ref{fig1}, the  values of $\psi$ at $\kappa = 4 \ko$ are
$\psi_1 = 83.25^\circ$ (on thin solid curve), $\psi_2= 167.63^\circ$
(on thin dashed curve),
 $\psi_3 = 39.78^\circ$
(on thin dashed  curve),  and $\psi_4 = 160.09^\circ$ (on thin solid
curve).

\begin{figure}[!ht]
\centering \psfull  \epsfig{file=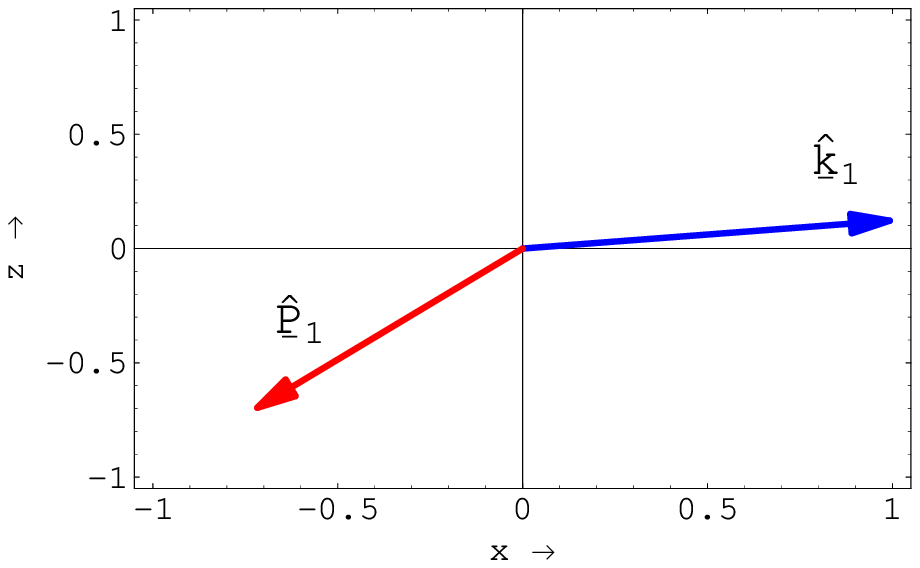,width=3.0in} \hfill
 \epsfig{file=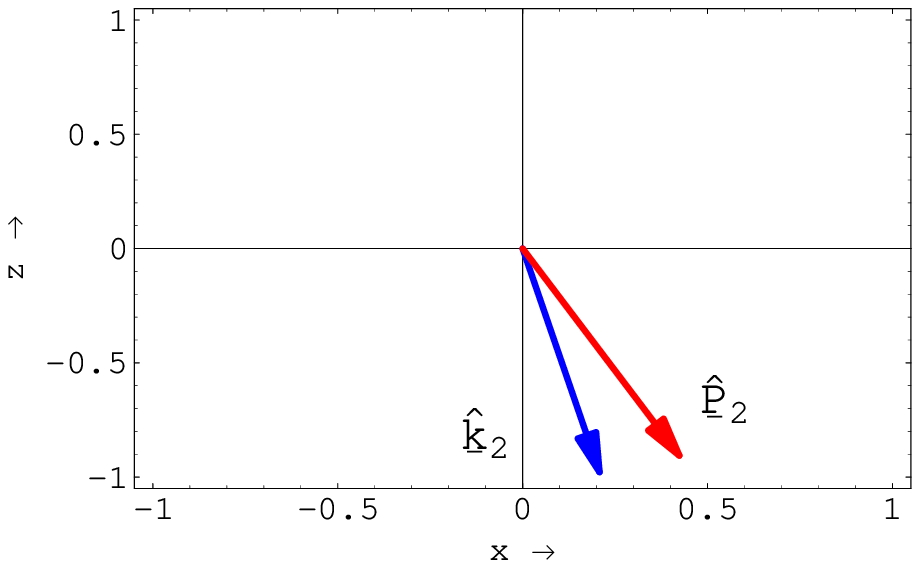,width=3.0in}\\
  \epsfig{file=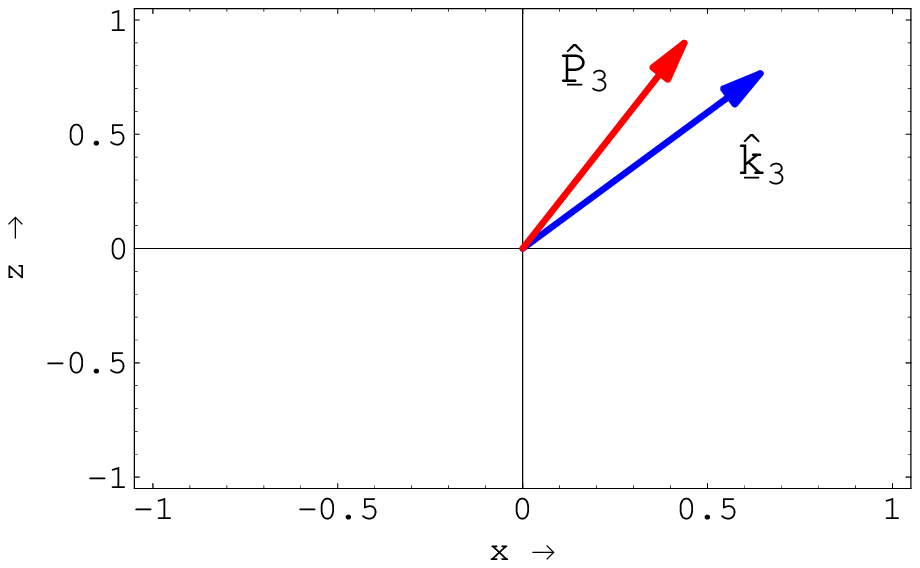,width=3.0in}
\hfill
 \epsfig{file=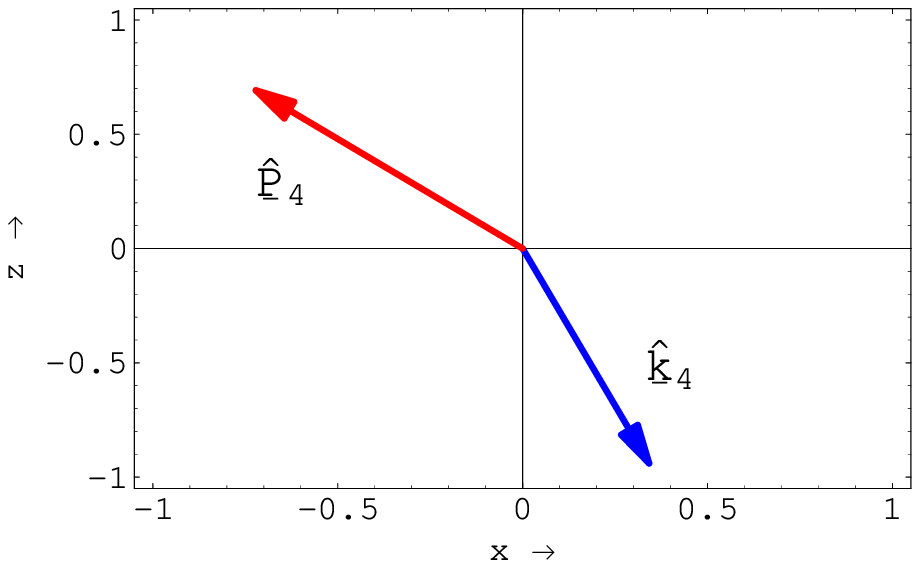,width=3.0in}
  \caption{\label{fig2}
  The orientations of the unit vectors $\hat{\#k}_j = \#k_j / k_j$ and $\hat{\#P}_j = \#P_j / P_j$ ($j \in \les 1,4 \ris $)
in the $xz$ plane for $\kappa = 4 \ko$, $\eps = 3 $, $\eps_g =
1$, $\eps_z = 2.5  $, $\xi = 10$, $\xi_g = 6 $, $\xi_z = 7  $, $\mu
= 2 $, $\mu_g = 0.5 $ and $\mu_z = 1.5 $.
 }
\end{figure}

The orientations of the  wavevectors $\#k_j$ and the time--averaged
Poynting vectors $\#P_j$  at $\kappa = 4 \ko$ in the FCM are
presented  in Figure~\ref{fig2}. This figure provides confirmation that the wavevectors
with orientation angles $\psi_{1,2}$ are associated with energy flow
towards the boundary  $z=0$, whereas the other two wavevectors are
associated with energy flow away from that boundary. In addition, the
 plane waves indexed $1$ and $4$ have negative phase velocity, while the
plane waves indexed $2$ and $3$ have positive phase velocity.

Now, let us consider further the reflection of the incident plane wave indexed
$1$
 by letting $A_1=1$ and $A_2 = 0$. This case
 results in a positively reflected plane wave with
wavevector orientation angle $\psi_4$ and a negatively reflected
plane wave with wavevector orientation angle $\psi_3$. The
positively reflected plane wave has negative phase velocity whereas
the negatively reflected plane wave has positive phase velocity.

Next, we let $A_1=0$ and $A_2=1$ in order for the plane wave
indexed $2$ to be incident on the boundary $z=0$. In this case, the
reflected plane wave with wavevector orientation angle $\psi_3$ is
positively reflected and has positive phase velocity, whereas the
plane wave indexed $4$ is negatively reflected and has negative
phase velocity.

In order to check that the reflected plane waves described in
Figure~\ref{fig2} have nonzero amplitude, the system of algebraic
equations \r{ae1}--\r{ae4} was solved at $\kappa = 4 \ko$. With  $ \#E_{\,0j} \. \#E^*_{\,0j}  =1 $ ($j \in
\les 1,4\ris$), we
obtained: $A_3 = 0.006 - 0.079 i$, $A_4 = -0.892 - 0.372 i$, $B_s =
0.078 - 0.306i$, and $B_p = -0.253 - 0.036 i$ when $A_1 = 1$ and
$A_2 = 0$; and $A_3 = -0.921 - 0.477 i$, $A_4 = 0.073 - 0.071 i$,
$B_s = 0.106 - 0.380i$, and $B_p = 0.329 + 0.070 i$ when $A_1 = 0$
and $A_2 = 1$. Therefore, the negatively reflected waves do indeed
have nonzero (but small)  amplitude. We note that  the refracted plane waves (in
the vacuous region $z < 0$) are evanescent, since $\alpha = i \ko
\sqrt{15}$ at $\kappa = 4 \ko$.

The amplitudes of the negatively reflected plane waves increase
if the vacuous region $z < 0$ is filled by a perfect electric
conductor. Then, we must set $B_s = B_p = 0$ and the  algebraic
equations \r{ae1} and \r{ae2} have to be solved to find $A_3$ and $A_4$.
At $\kappa = 4 \ko$ with  $ \#E_{\,0j} \. \#E^*_{\,0j}  =1 $ ($j \in
\les 1,4\ris$), we found: $A_3 = -0.884$ and $A_4 = -0.089$ when
$A_1 = 1$ and $A_2 = 0$; and $A_3 = 0.102$ and $A_4 = -1.139$ when
$A_1 = 0$ and $A_2 = 1$.

\newpage
\noindent{\bf 5. CONCLUDING REMARKS}

The four independent wavenumbers associated with planewave
propagation in a FCM can give rise to a host of complex
electromagnetic behavior. With a nondissipative example, we have shown
that an incident plane wave with positive phase velocity
can result in a negatively reflected plane wave with negative phase
velocity, as well as a positively reflected plane wave with
positive phase velocity. Also, an incident plane wave with negative
phase velocity can result in a negatively reflected plane wave with
positive phase velocity, as well as a positively reflected plane
wave with negative phase velocity. Thus, negative reflection
is characterized by a reversal in the co/contra directionality
of the wavevector and the time--averaged Poynting vector.
These negative reflection
characteristics add to an already rich palette of electromagnetic
responses supported by linear bianisotropic mediums
\c{PiO}.

\vspace{10mm}

\noindent{\bf Acknowledgments:} During part of this study TGM was
supported by a \emph{Royal Society of Edinburgh/Scottish Executive
Support Research Fellowship}. AL thanks the Charles Grover Binder
Endowment at Penn State for partial support.

\newpage

\newpage

\end{document}